\documentclass[amsmath,amssymb,aps,prb,reprint,twocolumn,unsortedaddress]{revtex4-2}
\usepackage{bm}
\usepackage[rgb]{xcolor}
\usepackage{graphicx}
\usepackage{physics}
\usepackage{comment}
\usepackage{enumitem}
\usepackage{todonotes}
\usepackage{hyperref}
\usepackage[capitalize]{cleveref}
\usepackage{dcolumn}
\usepackage[caption=false]{subfig}
\usepackage{verbatim}

\begin{document}

\title{A bilayer cellular Potts model of epithelial docking}

\author{Troy Singletary}
\email{troy.singletary@unco.edu}
\affiliation{Department of Physics and Astronomy, University of Northern Colorado, Greeley, CO 80639}

\author{Andrea James}
\email{andrea.james@unco.edu}
\affiliation{Department of Biological Sciences, University of Northern Colorado, Greeley, CO 80639}

\author{Tyler A. Engstrom}
\email{tyler.engstrom@unco.edu} 
\affiliation{Department of Physics and Astronomy, University of Northern Colorado, Greeley, CO 80639}

\date{\today}

\begin{abstract}
 Fusion of two epithelial cell sheets brought together in a bilayer configuration is a common step in animal morphogenesis, yet, in contrast to other epithelial fusion processes such as wound healing in a monolayer of cells, it has not been a strong focus of modeling efforts. Here we consider a preliminary stage of bilayer fusion, recently termed ``docking”~\cite{cote2022wont}. In multiple instances of docking that span apical and basal varieties, cells appear to have a tendency to remodel so as to co-localize their bilateral junctions (match their edges) across the bilayer. Motivated by this observation, we introduce a bilayer cellular Potts model that couples two standard 2D area- and perimeter-elasticity models via short-range, out-of-plane interactions between cell edges. The new coupling involves a single adjustable parameter that minimally models the combined effect of dynamic cytoskeletal protrusions, cadherins, and other potential edge-associated adhesion molecules. Our model predicts that bilayer edge matching is maximized when the two monolayers are in their fluid-like regimes (average cell shape index $\gtrsim4.6$ in our hexagonal lattice implementation), and when the bilayer coupling strength strikes a balance between in-plane and out-of-plane energy scales. At higher bilayer coupling strengths, the system tends to get stuck in metastable states with sub-optimal edge matching. Exploration of the mechanisms of edge matching reveals that pairs and quadruplets of coordinated T1 transitions play a particularly important role. We also find numerous examples of emergent features we term ``domain walls” — branching or unbranching curves that cross no matched edges, but that separate regions of nearly complete matching. These domain walls can be both system spanning and long lived. Finally, we extend our model to crudely account for bending of the two epithelial sheets, and study the distributions of docking front speeds that result.
\end{abstract}

\maketitle


\section{\label{sec:level1}Introduction }


A versatile and topology-altering step in morphogenesis is a face-to-face ``docking'' and subsequent fusion of two epithelial cell sheets that have been brought into proximity by growth, folding, and/or other large-scale tissue movements~\cite{andrew2010morphogenesis, ray2012mechanisms, cote2022wont, ishihara2023topological}. By this mechanism, through-gut body plans may be established, sheets may pinch off into tubes, tubes may compartmentalize into more tubes, and bridges may be created between branching structures. The sheets in question may dock either at their apical surfaces, e.g., in neural tube closure~\cite{nikolopoulou2017neural, mole2020integrin, hashimoto2015sequential, hashimoto2019differential, macgowan2025fold}, foregut compartmentalization~\cite{billmyre2015one, edwards2025disrupted}, and palate fusion~\cite{teng2026actomyosin}, or, following basement membrane breakdown, at their basal surfaces, e.g., in parabronchial fusion~\cite{palmer2020fusion} and nephrogenesis~\cite{kao2012invasion}. Optic fissure closure is a well-studied and interesting hybrid case in which the docking occurs initially between basal surfaces, but the cell polarity then reverses and apical-apical fusion results~\cite{james2016hyaloid, gestri2018cell, bernstein2018cellular, eckert2020vivo, chan2021closing}. Similar apicobasal polarity changes occur in other epithelial docking events where they facilitate the cell neighbor exchange that results in a continuous lumen~\cite{portereiko2001early, cote2025reciprocal}.

\begin{figure}[t]
\centering
\includegraphics[width=0.9\linewidth]{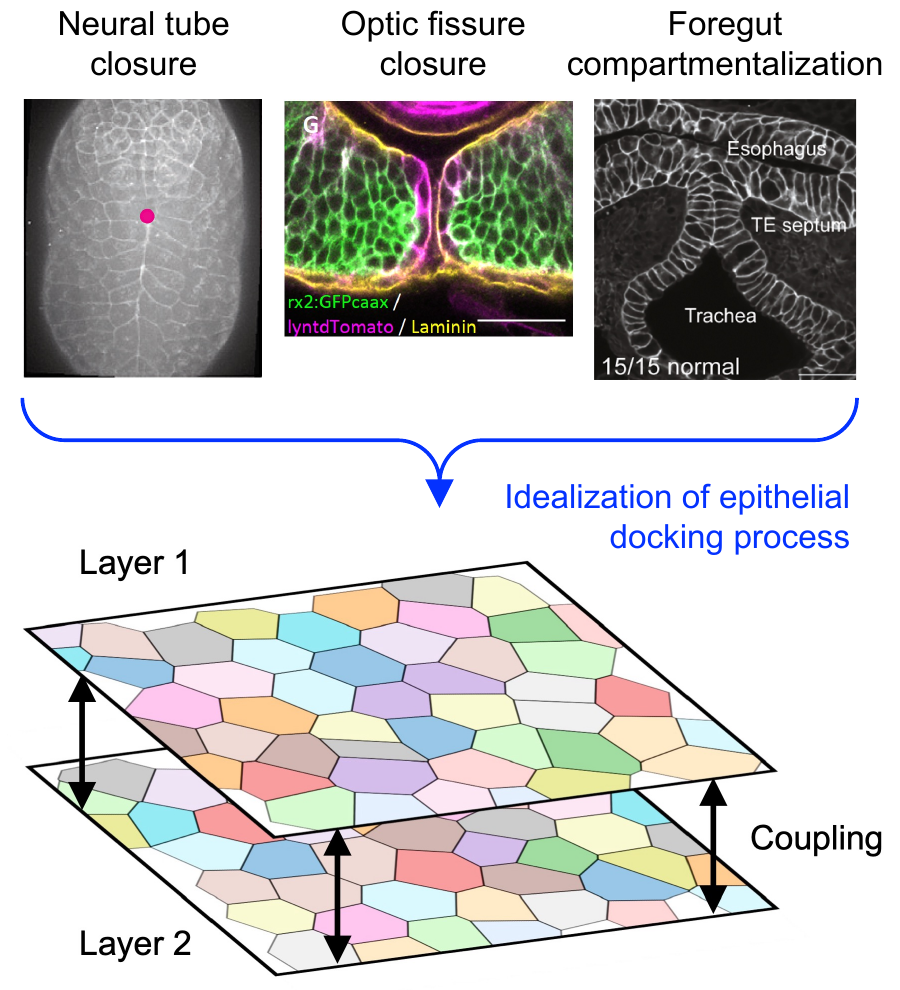}
\caption{Multiple forms of epithelial docking may be amenable to minimal modeling with a bilayer system of 2D cellular models. In the bilayer cellular Potts model (BCPM) we introduce here, the coupling is between cell edges, but other kinds of coupling (e.g., between cell vertices, centers, or a combination) are also possible. The top left image of neural tube closure is reprinted from Fig. 6B ($T=3600$s) in Ref.~\cite{hashimoto2015sequential} with permission from Elsevier; the top center image of optic fissure closure is reprinted from Fig. 2G in Ref.~\cite{eckert2020vivo} under the Creative Commons CC-BY 4.0 license; the top right image of foregut compartmentalization is reprinted from from Fig. 4B (control) in Ref.~\cite{edwards2025disrupted} under the Creative Commons CC-BY-NC-ND 4.0 license. Note the high degree of cell edge matching in each of these experimental images.}
\label{fig:BCPMschematic}
\end{figure}

Many instances of epithelial docking involve dynamic cytoskeletal protrusions that mediate adhesion of the bilayer~\cite{cote2022wont}. Much about the specific adhesion molecules and the specialized cell types that express them is unknown, but it is clear that at least in some instances of docking (e.g., those shown in Fig.~\ref{fig:BCPMschematic}) they promote the pairing-up of cells and near-mirroring of cell shapes across the bilayer so that the coordinated polarity rotations necessary for fusion can occur. One possible and simple mechanism of mirroring may be the intercalation or interdigitation of the cytoskeletal protrusions~\cite{ray2012mechanisms, palmer2018epithelial}: the more regular the intercalation, the stronger the adhesion that may be expected to result. Further evidence that dynamic cytoskeletal protrusions facilitate cell shape matching and fusion comes from studies of \emph{Drosophila} dorsal closure~\cite{jacinto2000dynamic, jacinto2001mechanisms}, however, this process, like wound healing, is geometrically distinct from docking as it involves epithelial sheets that sweep over a substrate and fuse edge-to-edge. Cross-bilayer interactions between existing adherens junctions~\cite{nikolopoulou2017neural} and the formation of new junctions between apposed cell sheets~\cite{teng2026actomyosin, popkova2024mechanical} likely also contribute to cell shape matching.

Errors in epithelial docking and fusion processes can lead to a variety of birth defects including colobomas (improper choroid fissure closure that accounts for up to 11\% of reports of childhood blindness worldwide~\cite{gregory-evans2004ocular}), spina bifida and anencephaly (neural tube defects that annually affect 300,000 newborns worldwide~\cite{kondo2009neural}), and cleft lip and/or palate (affecting one in 700 live births worldwide~\cite{dixon2011cleft}). Quantitative biophysical modeling could provide a better understanding of the cellular mechanisms underlying epithelial docking, potentially aiding the development of medical treatments for these disorders.

While three-dimensional (3D) cellular models such as vertex~\cite{okuda2015three, zhang2022topologically, sarkar2026reentrant}, Voronoi~\cite{merkel2018geometrically}, lattice-based~\cite{swat2012multi}, deformable cell~\cite{runser2024simucell3D}, and active foam~\cite{ongenae2025active} models would be attractive starting points from which to build a realistic model of epithelial docking, it is interesting to ask whether a more minimal model can be constructed using a bilayered system of two-dimensional (2D) cellular models, each of which represents a cross section of an epithelial tissue taken parallel to the apical surface. A schematic of this is shown in Fig.~\ref{fig:BCPMschematic}. Tissue fusion in 2D cellular models has been extensively studied in the context of in-plane wound healing, i.e., the closure of a hole in a monolayer of cells via unjamming and actin cabling~\cite{tetley2019tissue, noppe2015modelling, almada2026physical}, as well as directional zippering as in \emph{Drosophila} dorsal closure~\cite{godeau2025transient}. However, to our knowledge, these models have not been harnessed to study \emph{out-of-plane} fusion such as occurs in connection with epithelial docking. 2D vertex models~\cite{nagai2001dynamic, fletcher2014vertex, bi2015density} would be a natural choice if the bilayer coupling acted between cell vertices, yet the cytoskeletal protrusions that mediate docking (as well as cadherins that are likely involved in the subsequent fusion~\cite{chen2012cadherin, nikolopoulou2017neural, teng2026actomyosin, popkova2024mechanical}) are not confined to cell vertices. Hence, we propose a simple, short-range bilayer coupling between cell \emph{edges}, which represent the cytoskeleton in 2D cellular models. We implement this new coupling between two otherwise standard cellular Potts models (CPMs), also known as Glazier-Graner-Hogeweg models~\cite{graner1992simulation, glazier1993simulation, hogeweg2000evolving}; the underlying hexagonal lattices of the CPMs always stay in registry and make it straightforward to determine whether a cell edge in layer 1 is co-localized (and can interact with) with a cell edge in layer 2. 

The resulting bilayer cellular Potts model (BCPM) exhibits several interesting behaviors owing to the interplay between the solid-like to fluid-like transition within the constituent CPM monolayers~\cite{chiang2016glass, sadhukan2021theory, devanny2023signatures, nemati2024cellular}, and the new bilayer coupling. Among these behaviors is a parametric sweet spot for docking where the monolayers are both in the fluid-like regime and the energy scales for in-plane and out-of-plane cell edge adhesion are balanced. We also see domain-wall-like structures indicative of geometric frustration between fused regions; these emergent defects have no counterpart in purely 2D cellular models. Finally, we show how the BCPM can be extended in a simple way to account for bending energy, and how this extension gives rise to a docking front that propagates with a speed distribution strongly dependent on model parameters. 

%
%
%
%
%
%


\section{\label{sec:level1}Modeling }


\subsection{\label{sec:level2} Standard cellular Potts model }

The cellular Potts Model (CPM) is a lattice spin model that uses groups of like spins, and their Wigner-Seitz cells as pixels, to represent biological cells. The spin value at lattice site $i$ is called the cell index $\sigma_i=1\dots N$, with $\sigma_i=0$ sometimes used to represent a substrate. This model describes cellular tissue dynamics under the influence of area elasticity and cell-cell adhesion using a Hamiltonian~\cite{graner1992simulation, glazier1993simulation, nemati2024cellular}  

\begin{eqnarray}
H = \lambda_{A}\sum_\sigma \left(A_{\sigma} - A_{0} \right )^2 + J \sum_{<i, j>} \left[ 1 - \delta_{\sigma_{i}, \sigma_{j}}\right].
\label{eq:Classic_CPM_Hamiltonian}
\end{eqnarray}
Here $A_{\sigma}$ is the area of cell $\sigma$, $A_{0}$ is the target area of each cell, $\lambda_{A}$ is proportional to the cell bulk modulus, $\delta_{\alpha, \beta}$ is the Kronecker delta, $J$ is the adhesion energy per pixel edge between cells (taken to be constant for simplicity), and $<i,j>$ denotes a nearest-neighbor pair of spins. 

The system is evolved via a Metropolis-Hastings Monte-Carlo algorithm where a move consists of a randomly chosen lattice site that has its cell index copied into a randomly chosen neighboring site. The new microstate corresponds to a change in energy $\Delta H$ and is accepted with probability

\begin{equation}
P(\Delta H) =
\left \{  
\begin{array}{lr}     
1, &  \text{if } \Delta H < 0 \\
e^{-\Delta H /T}, & \text{if } \Delta H  \ge 0 
\end{array}
\right \}	,
\label{eq:Monte_Carlo_Prob}
\end{equation}
where the effective temperature $T$ sets the rate of cell membrane fluctuations. Time is measured in units of Monte-Carlo steps (MCS) where one MCS is defined as a number of copy attempts (moves) equal to the number of lattice sites in the system.

\subsection{\label{sec:level2}Bilayer cellular Potts model}

Here, we introduce a novel bilayer cellular Potts model (BCPM) that couples two planar 2D CPMs as depicted in Fig~\ref{fig:BCPMschematic}. In constructing the BCPM, we first modify Eq.~\eqref{eq:Classic_CPM_Hamiltonian} to the area- and perimenter-elasticity form
\begin{equation}
H_{\textrm{monolayer}} = \sum_\sigma \left [ \lambda_{A}\left( A_{\sigma} - A_{0}\right)^2 
+ \lambda_{P}\left( P_{\sigma} - P_{0} \right)^2 \right ].
\label{eq:Monolayer_Hamiltonian}
\end{equation}
The standard CPM adhesion term (proportional to the total cell perimeter) has been absorbed into a polynomial perimeter term in which $P_\sigma$ is the perimeter of cell $\sigma$ and $P_0$ is the target perimeter of each cell. Qualitatively similar to vertex models~\cite{nagai2001dynamic, fletcher2014vertex, bi2015density}, $\lambda_P$ here can be regarded as a spring constant for the active, actomyosin-containing cell perimeter, $2\lambda_PP_0$ is the cadherin-mediated adhesion energy per unit length between cells, and a dimensionless ``cell shape index'' $p_0=P_0/\sqrt{A_0}$ controls whether the monolayer CPM system is in a more ordered, solid-like state (low $p_0$) or a more disordered, fluid-like state (high $p_0$)~\cite{noppe2015modelling, sadhukan2021theory, devanny2023signatures}. However, the nature of this transition in CPMs can vary from glassy to first-order depending on implementation details such as lattice type and a linear~\cite{chiang2016glass, nemati2024cellular} versus quadratic~\cite{noppe2015modelling, sadhukan2021theory, devanny2023signatures} perimeter term, and recent work~\cite{sadhukan2021theory, nemati2024cellular} cautions against interpreting the cell-averaged shape index $\langle P_{\sigma}/\sqrt{A_{\sigma}}\rangle$ as a structural order parameter as it is in vertex models~\cite{bi2015density, bi2016motility}. While these details are clearly important, the exact nature of the monolayer transition will not be a main focus of the current work.

\begin{figure}[t]
\centering
\includegraphics[width=0.9\linewidth]{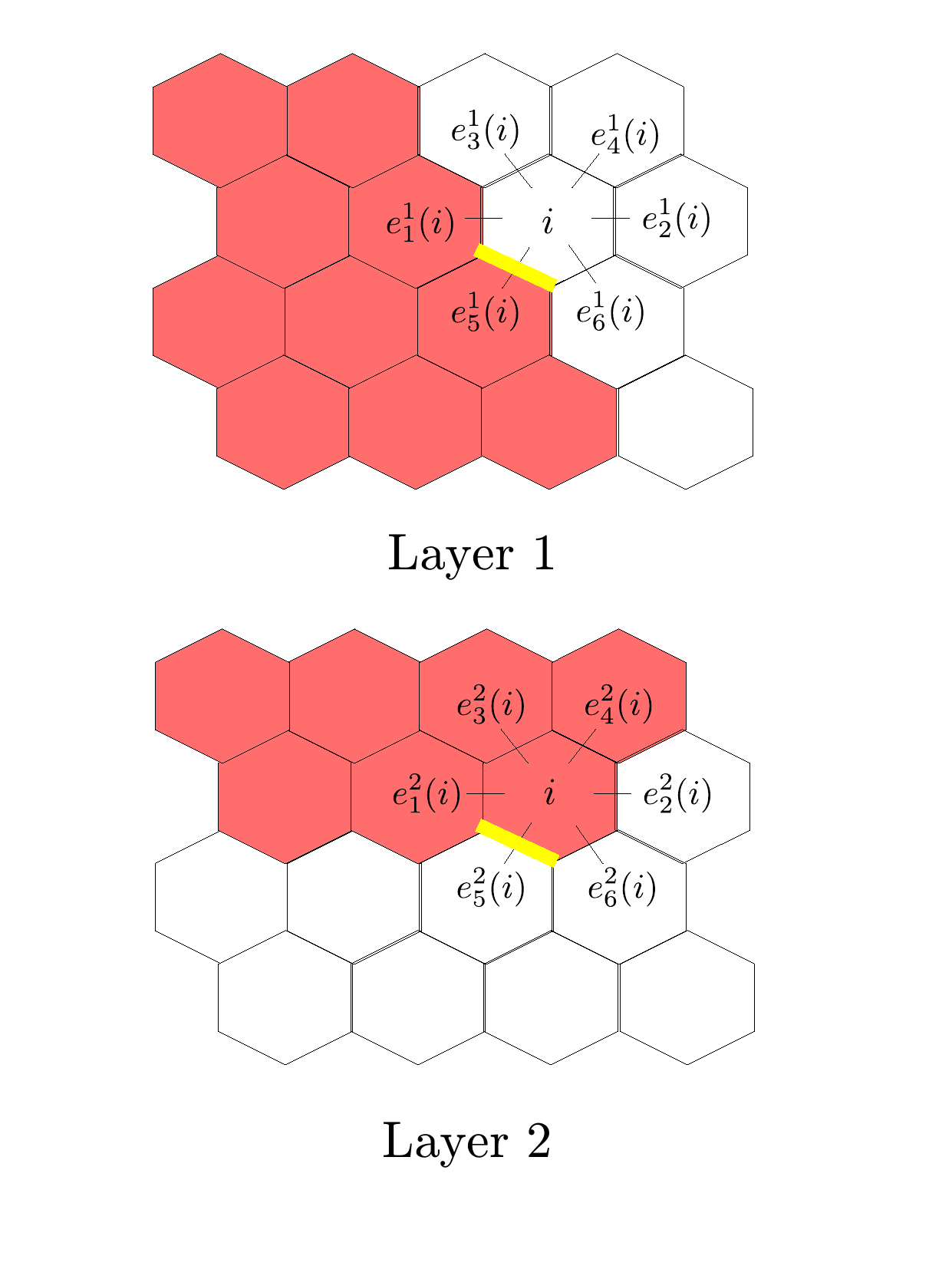}
\caption{An example of cell edge matching in the BCPM, where the red and white Wigner-Seitz cells (pixels) belong to different epithelial cells. Consider lattice site $i$ which has the same $x$-$y$ coordinates in each layer. Site $i$ borders multiple cell edge segments in each layer, but edge segment number 5 (in yellow highlight) is the only matching edge, thus the only edge that contributes to Eq.~\eqref{eq:Coupling_Hamiltonian}}.
\label{fig:Edgematch_definition}
\end{figure}

Next, we couple together two monolayer CPMs such that
\begin{equation}
H_{\textrm{bilayer}} = H_{\textrm{layer 1}} + H_{\textrm{layer 2}} + H_{\textrm{coupling}}.
\label{eq:Bilayer_Hamiltonian}
\end{equation}
Various forms of $H_{\textrm{coupling}}$ are possible, for example, the cell vertices in layer 1 could interact with the cell vertices in layer 2, or the cell centers could interact, or a combination. Here, however, we implement a coupling between the cell edges in each layer, which is motivated by the above-discussed molecular mechanisms that appear to be involved in epithelial docking, as well as experimental observations of cell shape matching (e.g., in Fig.~\ref{fig:BCPMschematic}). The form of our coupling introduces a single new adjustable parameter $\lambda_B>0$ and is given by
\begin{equation}
H_{\textrm{coupling}} = -2\lambda_B \sum_{\langle i,j \rangle} e_j^{1}(i) e_j^{2}(i). 
\label{eq:Coupling_Hamiltonian}
\end{equation}
The function $e_j(i)=1-\delta_{\sigma_{i}, \sigma_{j}}$ is the same as that in Eq.~\eqref{eq:Classic_CPM_Hamiltonian} and equals 0 if nearest-neighbor lattice sites $i$ and $j$ don't share a cell edge; equals 1 if they do. The superscript indicates which layer this function is being applied to. Thus, this coupling reduces the energy by $2\lambda_B$ for every pixel edge that forms part of an epithelial cell edge in both layers, as shown in Fig.~\ref{fig:Edgematch_definition}. While Eq.~\eqref{eq:Coupling_Hamiltonian} does not explicitly reference the interlayer distance, we have in mind a distance within the finite interaction range of the adhesion molecules.  

\subsection{\label{sec:level2}Simulation Details}

The standard system studied here features two monolayers consisting of $102 \times 102$ site hexagonal lattices with periodic boundary conditions. Each lattice is independently initialized into random configurations of confluent cells. A relatively small number of cells (10-40) are used, based on the number of cells involved in a variety of epithelial docking processes (Fig.~\ref{fig:BCPMschematic}). Following the convention used in CC3D~\cite{swat2012multi}, the hexagonal lattice scaling is set such that the Wigner-Seitz cell area $a$ is $1$, from which the nearest neighbor distance becomes $L=\sqrt{\frac{2\sqrt{3}}3} \simeq 1.07$, the site interface length (yellow line segment in Fig.~\ref{fig:Edgematch_definition}) becomes $s =\frac{\sqrt{2\sqrt{3}}}{3} \simeq 0.62$, and the domain size is $L_{x} = \frac{205}{2}\sqrt{\frac{2\sqrt{3}}3} \simeq 110.1 $, $L_{y} = 102\frac{\sqrt{3}}{2}\sqrt{\frac{2\sqrt{3}}3} \simeq 94.9$. Equations \eqref{eq:Monolayer_Hamiltonian}-\eqref{eq:Coupling_Hamiltonian}  are implemented in a form that explicitly shows the perimeter definition and its relationship to the new bilayer coupling: 
\begin{align}
H = \sum_{\sigma} \Bigg{[}&\lambda_{A} \left( A_0 - \sum_{i \in \sigma} a \right)^2 \nonumber \\ 
+&\lambda_{P} \left( P_0 - s\sum_{\substack{\langle i,j\rangle \\ i \in \sigma}}  e_{j}^{1,2}(i) \right)^2 \nonumber\\
-&\frac{\lambda_{B}}{2} \sum_{\substack{\langle i,j\rangle \\ i \in \sigma}}e_{j}^{1}(i) e_{j}^{2}(i) \Bigg{]}.
\label{eq:Bilayer_Hamiltonian_sites}
\end{align}
Here, the sum over cells includes all cells in the bilayer. Because each cell edge is shared between two cells in the same layer, and each matching edge is shared between both layers, the coefficient of the coupling term here is 4 times smaller than that in Eq.~\eqref{eq:Coupling_Hamiltonian} to correct for this over-counting.

\begin{figure}[t]
\centering
\includegraphics[width=0.9\linewidth]{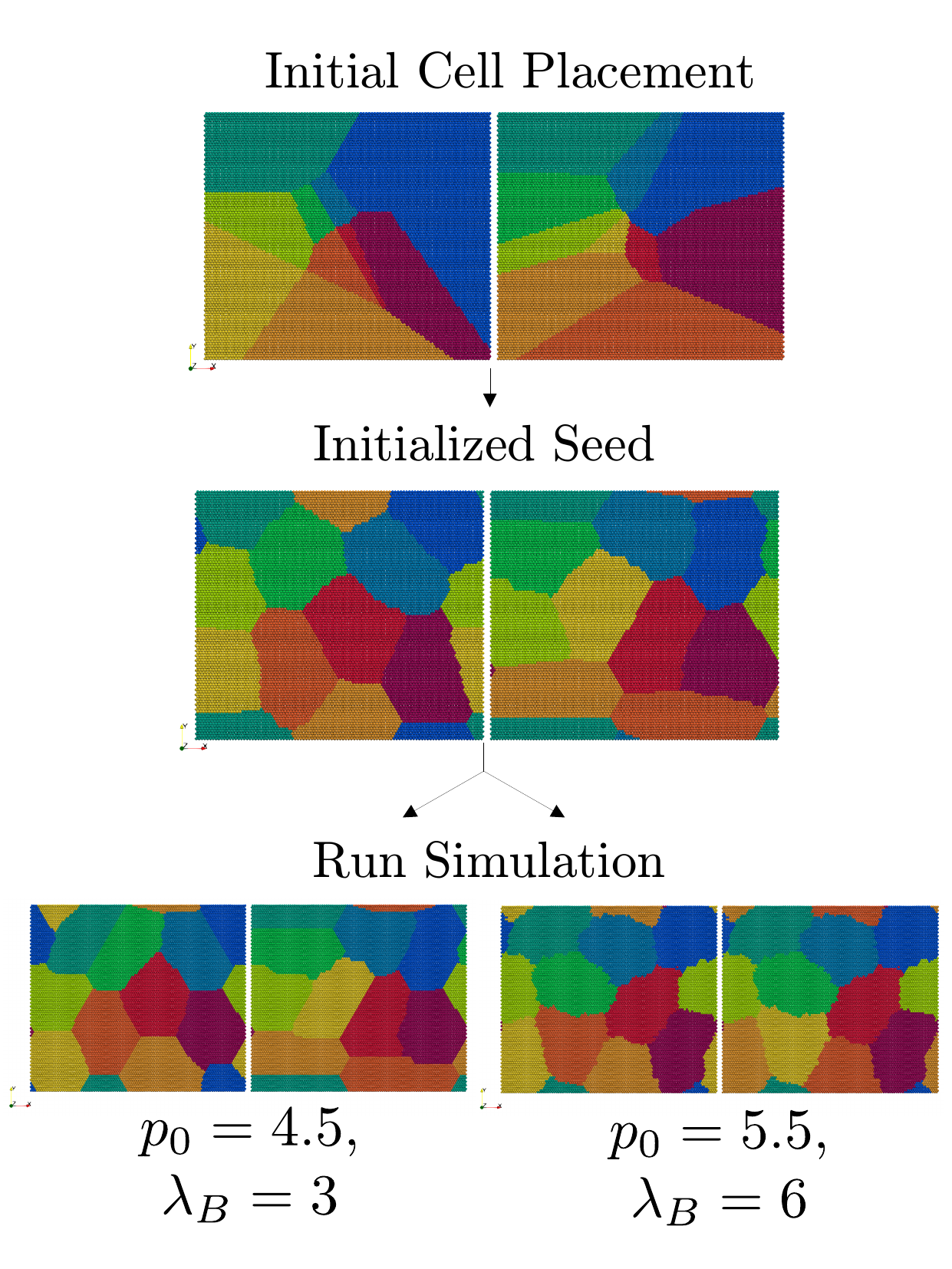}
\caption{Initial cell positions are generated by pseudo-randomly placing Voronoi points (here 10) in each layer and assigning cell indices $\sigma$ to lattice sites based on the Voronoi tiling. An initialization run is performed with $\lambda_{A} = 4, \lambda_{P}=1, \lambda_{B} = 0, p_{0} = 4.5$ for 500 MCS to allow each cell to reach area $A_{\sigma} \simeq A_{0}$, and the initialized configurations are denoted ``seeds" and used for simulation runs.}
\label{fig:Simulation_workflow}
\end{figure}

The system is initialized using a Voronoi tiling influenced setup where a number of Voronoi points equal to the number of cells $N$ in each layer are placed in the layer and sites are assigned to cells based on the Voronoi point with minimum distance. Picking Voronoi points completely at random runs the risk of two or more points being placed too close together and cells starting with small enough areas that they are unrealistically absorbed into other cells, so we apply a psuedo-random algorithm that maintains a certain minimum distance between the points~\cite{nemati2024cellular}. Voronoi points are determined by centering a circle of radius $L_x/3$ in each lattice and selecting $N$ equally spaced points on the circle with the $k^{th}$ point at coordinate $(x_{k}, y_{k}) = L_{x}/3( \cos{\frac{2 \pi k}{N}}, \sin{\frac{2 \pi k}{N} } )$. The $N$ points are all rotated by a random angle between $0$ and $2\pi$ and each point has its radial distance divided by a randomly chosen factor so it has a final radial distance $L_{x}/15 \le r_{k} \le L_{x}/3$.  Figure \ref{fig:Simulation_workflow} shows an example of this initialization scheme. These initially placed cells have a large variance in area so an equilibration run is performed using Eq.~\eqref{eq:Bilayer_Hamiltonian_sites} with $\lambda_{A} = 4$, $\lambda_{P} = 1$, $\lambda_{B}=0$, $T=10$, and $p_{0} = 4.5$ to allow each cell to reach area $A_{\sigma} \simeq A_{0}$ while maintaining realistic, e.g., unfragmented, shapes. 

In this work, we are primarily interested in studying possible interplay between the new bilayer coupling and the monolayer solid-like to fluid-like transition, so we focus on a minimal manipulation of model parameters expected to govern this interplay. We perform simulations that vary $0.5 \le \lambda_{B} \le 9.5$ and $4.0 \le p_{0} \le 6.0$ while fixing $T=10$, $A_0=1040$, $\lambda_{A} = \lambda_{P} = 1$. To quantify docking, we define an ``edge match ratio'' $M$ as the ratio of the matched cell perimeter to the average monolayer cell perimeter. Simulations are run for 5000 MCS, which allows almost every configuration studied to reach a steady-state value of $M$, which we denote $M_f$ and compute as the average of $M$ over the last 500 MCS. Unless otherwise noted, 20 distinct initializations (seeds) are used produce data for ensemble averages, and ensemble-averaged quantities are denoted with angle brackets, e.g., $\langle M_f \rangle$. In the following, time $t$ is used interchangeably with MCS. 
%
%
%

\section{\label{sec:level1}Results }

   
\subsection{Model validation at $\lambda_{B} = 0$}

\begin{figure}[t]
\centering
\includegraphics[width=\linewidth]{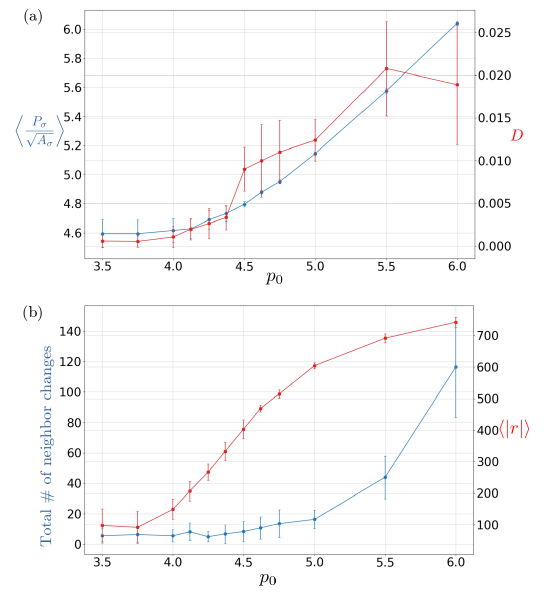}
\caption{Various monolayer ($\lambda_B=0$) properties as functions of target shape index $p_0$. The diffusion coefficient $D$ and observed shape index $\langle P_{\sigma}/\sqrt{A_{\sigma}} \rangle$ are shown in (a) while the distance traveled $\langle |r| \rangle$ and number of neighbor changes are shown in (b). Here, ensemble averages are over 10 different seeds.}
\label{fig:Diffusion_shape_distance_neighbor}
\end{figure}
\begin{figure}
\includegraphics[width=\linewidth]
{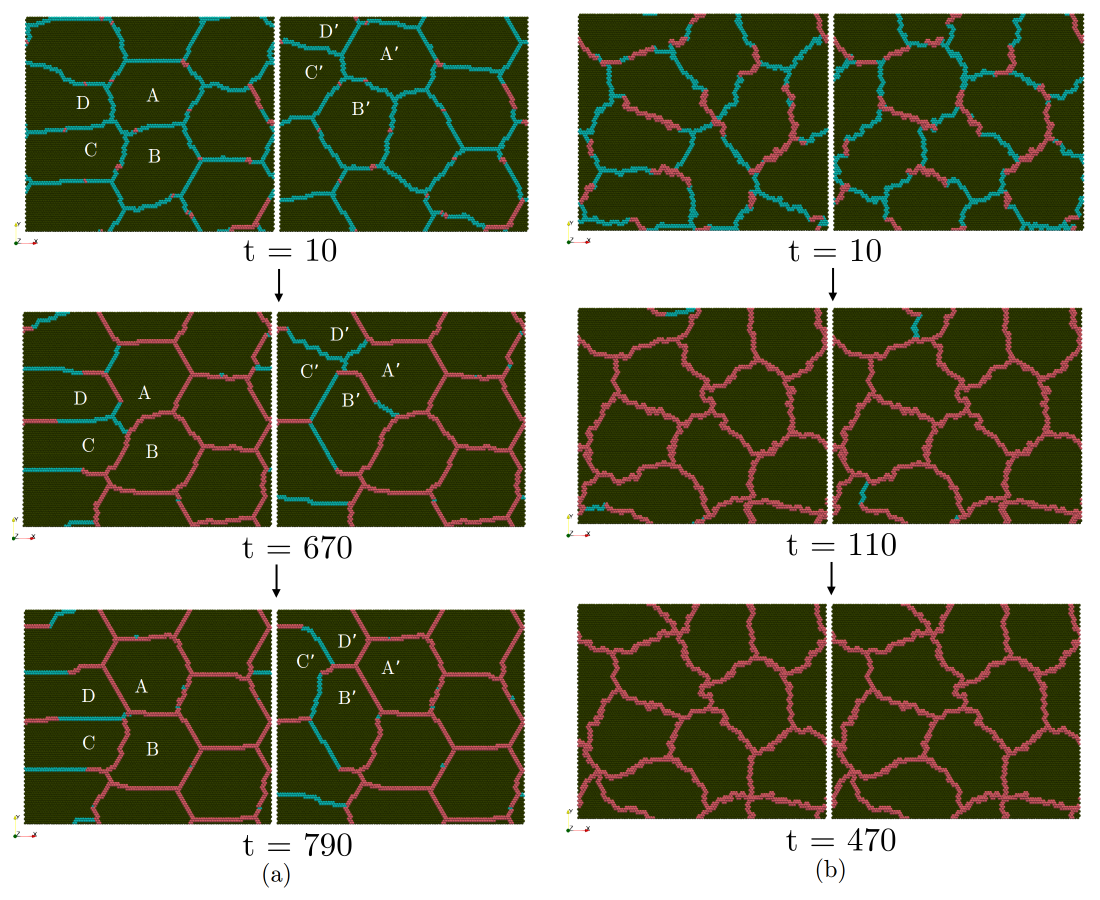}
\caption{Spatiotemporally coordinated T1 transitions are a primary mechanism of edge matching in the BCPM. Here and below, blue and red colors indicate unmatched and matched cell edges, respectively. Panel (a) shows a $p_{0} = 4.5$, $\lambda_{B} = 5$ system that undergoes simultaneous T1 transitions in each layer (involving the labeled cells). Panel (b) shows a $p_{0}=6$, $\lambda_{B} = 8.75$ system that undergoes four simultaneous T1 transitions (two in each layer), all of which arrest just after the transition point. Note also the rougher and less-hexagonal cell shapes in the larger $p_0$ system. Animations can be found in the Supplemental Materials~\cite{supMat}}. 
\label{fig:T1_and_rotation_Resolution}
\end{figure}

We initially set $\lambda_{B} = 0$ to see if our simulations can recover the known solid-like to fluid-like transition in a 2D CPM. As mentioned above, the nature of this transition depends strongly on model implementation details, so we focus on comparison with a similar hexagonal lattice CPM that uses a quadratic perimeter term in the Hamiltonian~\cite{noppe2015modelling}. In particular, we plot the observed average cell shape index $\langle P_\sigma/\sqrt{A_\sigma}\rangle$, the cell-center diffusion coefficient $D$ and the average distance traveled $\langle|r|\rangle$ (see details in Appendix), and the number of neighbor changes, all versus the target shape index $p_0$ and after $5000$ MCS. These data are shown in Fig.~\ref{fig:Diffusion_shape_distance_neighbor} and indicate a solid-like or ``hard'' regime (characterized by near-vanishing diffusivity and an approximately constant shape index) and a fluid-like or ``soft'' regime (much larger diffusivity and linearly increasing shape index) with the crossover occurring in the range $p_0=4-5$, consistent with Ref.~\cite{noppe2015modelling}. The observed saturation value $\langle P_\sigma/\sqrt{A_\sigma}\rangle\approx4.6$ is also similar to the values 4.8-4.9~\cite{chiang2016glass} and 4.11~\cite{sadhukan2021theory} reported in other CPM studies. Cell center-of-mass speed ranges from $0.03 \le c \le 0.15$ pixels/time in the range $4 \le p_{0} \le 6$, which is approximately the speed needed to travel the length of the system five times in a $5000$ MCS simulation for a cell moving in a straight line.

\subsection{\label{sec:level2}Edge matching over time}

\begin{figure*}
\centering
\includegraphics[width=\linewidth]{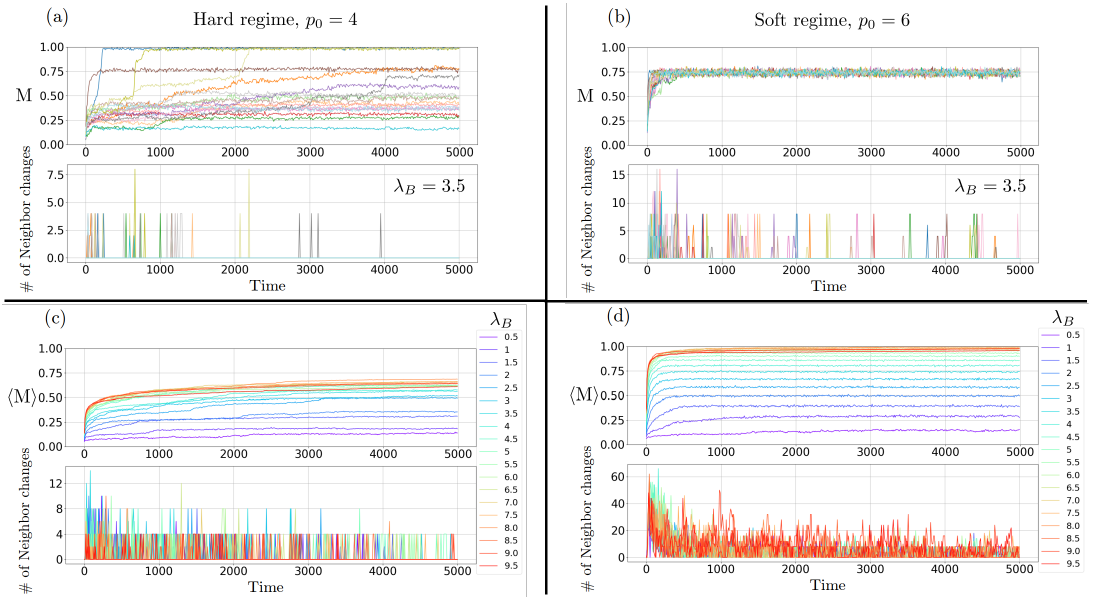}
\caption{
Edge match ratio $M$ vs time (top of each panel) and neighbor changes vs time (bottom of each panel). Panels in the top row show individual simulations with $\lambda_{B} = 3.5$ and (a) $p_{0} =4$ and (b) $p_{0}=6$. Panels in the bottom row show ensemble averages with variable $\lambda_B$ but otherwise the same parameters. In the hard regime, large neighbor change events appear correlated with large increases in $M$. }
\label{fig:Edgematch_vs_time}
\end{figure*}

 Figure~\ref{fig:T1_and_rotation_Resolution} shows typical simulation results for finite $\lambda_B$. T1 transitions that are coordinated within and/or between layers serve to increase the edge match ratio $M$ over time. Figure~\ref{fig:Edgematch_vs_time} plots the time evolution of $M$ at the lower and higher $p_{0}$ bounds in our study, with (a) and (b) giving results for individual simulations and (c) and (d) giving ensemble averages for different values of $\lambda_{B}$. At low $p_{0}$, cells are more rigid and the system takes longer to reach its saturated edge match ratio $M_f$. Individual simulations run with the same parameters can reach a wide range of $M_f$ values, occasionally including $1$ (complete matching). At high $p_{0}$, cells are more fluid and can better accommodate the rearrangements necessary for bilayer matching, and so the system reaches $M_f$ more quickly. Individual simulations run with the same parameters tend to reach a narrow range of $M_f$ values, which is controlled by $\lambda_B$.  
 To better understand the edge match results and identify mechanisms responsible for complete matching, we look at some associated system properties. For each cell, a neighbor list tracks adjacent cells, and a neighbor change is defined as a unit increment or decrement of the neighbor list. A T1 transition, for example, involves 4 neighbor changes, because two of the involved cells each gain a neighbor, and the other two each lose a neighbor. Cell rearrangements that involve fewer than 4 neighbor changes are also possible. For example, a 4-fold vertex can split into two 3-fold vertices and produce two neighbor changes, which can be viewed as half of a T1 transition. For sufficiently small system sizes, cells that lose contact in a T1 event can still remain neighbors via the periodic boundary conditions, with an example of this shown in the Appendix. In Fig.~\ref{fig:Edgematch_vs_time} we also plot the neighbor changes over time to examine whether they are correlated with increases in edge match ratio, and at low $p_{0}$ this appears to be the case: in panel (a), the simulations with large jumps in their edge match ratios undergo simultaneous rearrangements involving 4 or 8 neighbor changes (interpreted as 1 or 2 T1 transitions in the bilayer system). This is notable as neighbor changes are otherwise sparse in those simulations. In the hard regime, neighbor changes appear to cease after the system reaches its steady state edge match ratio. In contrast, the soft regime is characterized by frequent neighbor changes even after the edge matching reaches steady state.

\subsection{\label{sec:level2}Properties at steady-state matching}

\begin{figure*}[t]
\centering
\includegraphics[width=\linewidth]{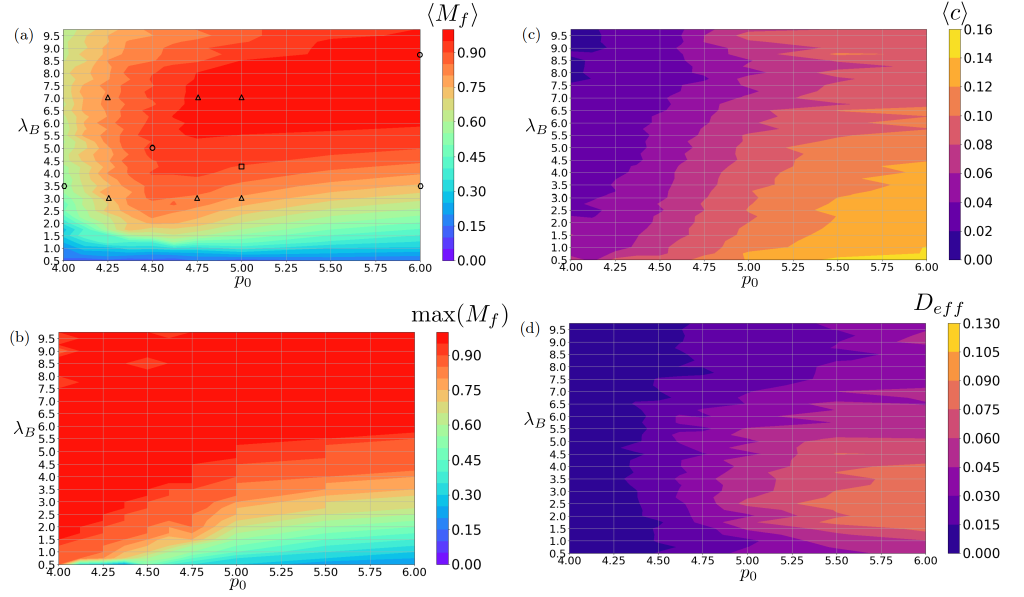}
\caption{BCPM regime maps parameterized by $\lambda_B$ and $p_0$. (a) Ensemble-averaged steady-state edge match ratio, where circles indicate the conditions used in Fig.~\ref{fig:T1_and_rotation_Resolution}, hexagons the conditions in Fig.~ \ref{fig:Edgematch_vs_time} (a) and (b), squares the condition in Fig.~\ref{fig:Domain_wall_30cells}, and triangles the conditions in Fig.~\ref{fig:CFC_edgematch_behavior}. (b) Maximum edge match ratio across the ensemble; (c) Average cell speed; (d) Effective diffusion coefficient.} 
\label{fig:Jamming_and_regime_map}
\end{figure*}
We regard $\langle M_f\rangle$ as a measure of the likelihood of successful epithelial docking under the given conditions. Figure~\ref{fig:Jamming_and_regime_map}(a) shows $\langle M_f\rangle$ as a function of $p_{0}$ and $\lambda_{B}$, and Fig.~\ref{fig:Jamming_and_regime_map}(b) shows the maximum value of $M_f$ across the ensemble for comparison. These two plots are nearly identical in the soft regime. The $\langle M_f\rangle$ plot exhibits a region with nearly complete matching that begins on the cusp of the soft regime at $p_0\approx4.7$ and occurs within an initially narrow band of $\lambda_B$ values centered on $\lambda_B\approx7$. The band gradually widens and trends to higher $\lambda_B$ values as $p_0$ increases. As a consistency check on these results, we also plot the ensemble-averaged cell speed $\langle c\rangle$ and the effective diffusion coefficient $D_{eff}$ (see details in Appendix) in Fig.~\ref{fig:Jamming_and_regime_map}(c) and (d), respectively. The former makes use of a control-theory-like characteristic time $\tau$ defined as the time taken for an individual simulation to reach $90\%$ of its $M_f$ minus the time taken to reach $10\%$ of that value. We then compute $\langle c\rangle$ as the average distance traveled during $\tau$ (see details in Appendix), divided by $\tau$. The results show that for $\lambda_B\lesssim7$, $\langle c\rangle$ depends strongly on $p_0$, increasing in the soft regime as expected. However, for $\lambda_B\gtrsim7$, $\langle c\rangle$ depends less strongly on $p_0$, consistent with the suppression of in-plane cell mobility due to large bilayer coupling.  For small $\lambda_B$, the iso-$D_{eff}$ contours indicate that the hard to soft transition shifts to smaller $p_0$ as $\lambda_B$ is increased. This direction of shift is notable in that it's the same as that induced by activity~\cite{bi2016motility}. At larger $\lambda_B$, the direction of shift changes. As a general trend, in the soft regime, $D_{eff}$ is larger where $\langle M_f\rangle$ is smaller, and vice versa.

\subsection{\label{sec:level2}Geometric frustration and domain walls}

\begin{figure}
\includegraphics[width=\linewidth]{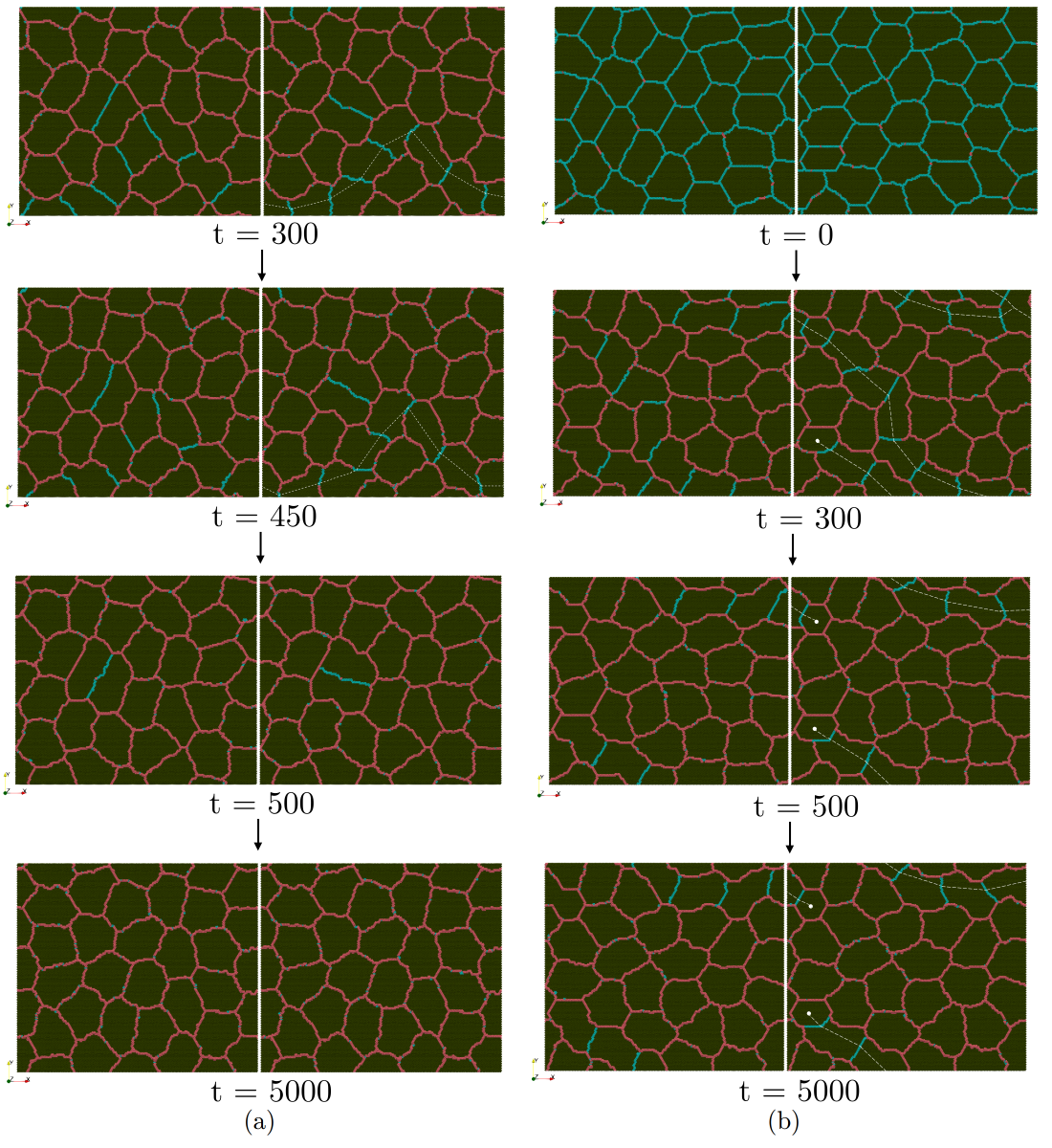}
\caption{Snapshots of edge matching in a larger system (30 cells, 180$\times$180 lattice sites in each layer) with $p_{0}=5$ and $\lambda_{B}=4.25$. In (a), the system quickly settles into a mostly matched configuration with a domain wall (dashed yellow line through blue edges) near the bottom. The unmatched edges of the domain wall then match via a coordinated, relative sliding, leaving a single pair of unmatched edges that are approximately perpendicular. Four spatio-temporally coordinated T1 transitions facilitate a relative rotation to match these edges. In (b), one long and branching domain wall partially resolves into a non-branching domain wall. This then persists to the end of the simulation. Animations can be found in the Supplemental Materials~\cite{supMat}}.
\label{fig:Domain_wall_30cells}
\end{figure}

A commonly observed type of defect in our simulations is a ``domain wall'' of unmatched edges that separates two completely matched tissue regions from each other. These domain walls can be either open or closed plane curves, and can range in length from several cell diameters to spanning the entire system. They can be transient defects that eventually resolve, or they can persist over the entire simulation time. Figure~\ref{fig:Domain_wall_30cells} shows several examples. A domain wall arises when two matched regions nucleate in different places, grow, and collide, leaving a group of cell edges in one layer that are roughly half-way in-between those in the other layer and present an obstacle to complete fusion. If and when they do fuse, they tend to do so all at once by a coordinated, relative sliding of the unmatched cell edges in the two layers (see movies in SM). This coordinating sliding is energetically favorable because it does not require cells to significantly change their areas or perimeters.

Sliding is often accompanied by small rotations of the unfused edges, and large rotations can also occur. The movie in SM shows an example of two perpendicular edges that fuse where they cross, and gradually torque each other into alignment (matching). Four coordinated T1 transitions (2 in each layer) facilitate this rotation, and lend further insight into neighbor change events that involve multiples of 4 neighbors, e.g., some of those in Fig.~\ref{fig:Edgematch_vs_time}. It's noteworthy that this edge rotation mechanism occurs naturally in our CPM, whereas in a vertex model setting where edges are straight, it would require a more complicated, distance dependent form of bilayer interaction. An interesting question is whether domain walls are a consequence of bilayer interaction over large length scales, and if they would be less frequent if the bilayer interaction were limited to smaller scales as with more realistic, curved epithelial sheets. This is the situation to which we now turn.

\subsection{\label{sec:level2}Crude Bending Model}

\begin{figure}[h]
\centering
\includegraphics[width=\linewidth]{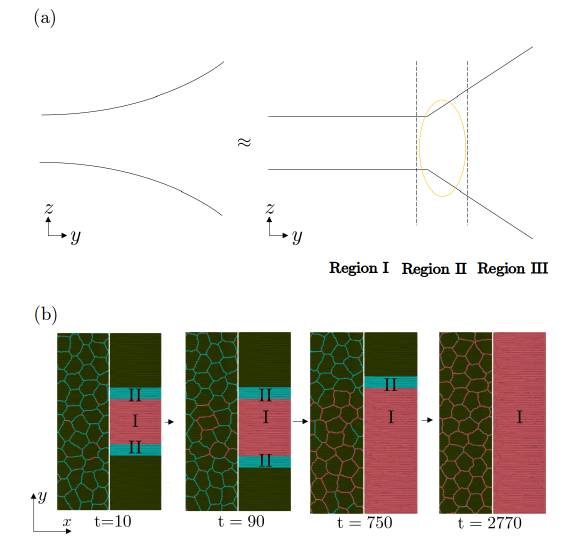}
\caption{(a) Approximation of bending as kinks in two otherwise flat surfaces. (b) Edge matching in the bending extension of the BCPM. In this simulation we take $K = 3$, $\epsilon= 0.5$, $p_{0}=4.75$ and increment $n$ whenever region II reaches $M = 0.4$. The $y>0$ and $y<0$ halves of the bilayer have the bending extension implemented independently, and the panels show $n=1,2,3,4$ for the top and bottom halves along with the respective times of those steps. Animation can be found in the Supplemental Materials~\cite{supMat}}.
\label{fig:Bending_Regions}
\end{figure}

So far we have explored epithelial docking in an idealized scenario where the two layers are parallel planar sheets coupled with a uniform $\lambda_{B}$. But realistic docking (such as the the examples shown in Fig.~\ref{fig:BCPMschematic}) involves curved tissue layers, and a more complete model would incorporate bending energies of the layers along with a distance-dependent $\lambda_{B}$ capturing the finite interaction range of the relevant adhesion molecules. While such a model is beyond the scope of the current work, we show how our planar BCPM is amenable to simple modifications that allow it to crudely account for bending of the two layers. To do this, we partition each layer into parallel flat (I); bent (II); and non-parallel flat (III) regions, as shown in Fig.~\ref{fig:Bending_Regions}(a). $\lambda_B$ is treated as a spatially dependent piecewise function
\begin{equation}
\lambda_{B}(y) = 
    \begin{cases}
        K, & |y| \le nL  \text{ (region I)},  \\
        K(1-\epsilon), & nL < |y| \le (n+1)L  \text{ (region II)},   \\
        0, &   |y|> (n+1)L  \text{ (region III)}, \\
    \end{cases}
\label{Bilayer_Gradient}
\end{equation}
where $K$ is the full bilayer coupling strength and $K(1-\epsilon)$ with $0 \le \epsilon \le 1$ is an effective bilayer coupling strength that captures both the bending energy incurred in region II and the greater average separation between adhesion molecules in region II. Furthermore, $L$ is a length scale on the order of $P_{0}$ in which the ``bending" is confined, and $n$ is an integer step parameter that is used to advance regions I and II as fusion progresses (whenever region II reaches a predetermined threshold value of $M$). We continue to use periodic boundary conditions in these bending simulations. 
\begin{figure}[h]
\centering
\includegraphics[width=\linewidth]{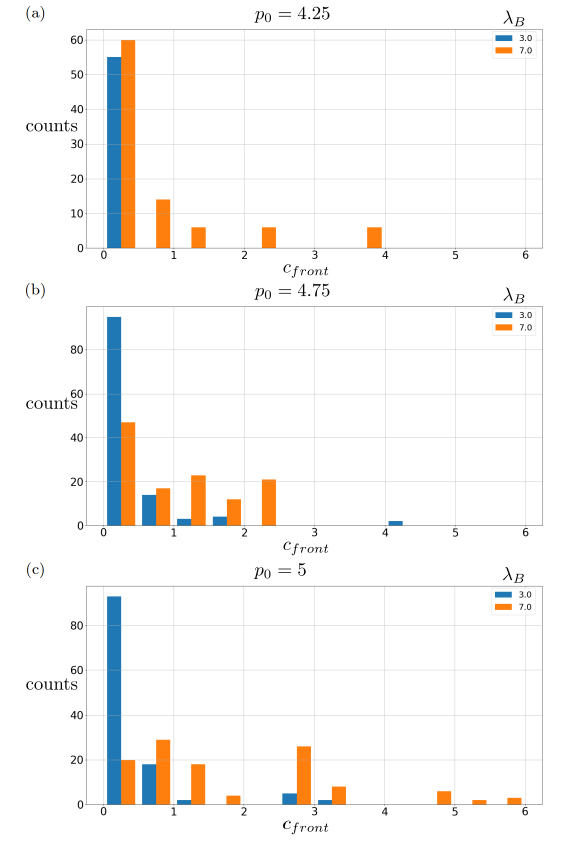}
\caption{Distributions of speeds $c_{front}$ at which region II advances, where $c_{front}$ is defined as the change in region I length divided by the change in time (MCS) between each advance. Ten different simulations are averaged over the conditions $p_{0} = 4.25, 4.75, 5.0$ (shown in different plots) and $\lambda_{B} = 3, 7$ (shown in blue and orange, respectively). }
\label{fig:CFC_edgematch_behavior}
\end{figure}

A sample bending simulation, in which the docking front (defined as the region I-II interface) advances over time, is shown in Fig.~\ref{fig:Bending_Regions}(b). The distribution of advancement speeds $c_{front}$ is shown in Fig.~\ref{fig:CFC_edgematch_behavior}. At low bilayer coupling strength ($\lambda_{B}=3$) the distribution is exponential-like with the tail lengthening as $p_{0}$ increases. At intermediate coupling strength ($\lambda_{B}=7$) however, the distribution changes from exponential-like at low $p_0$ to more uniform at high $p_0$. These results suggest the biological system may be able to tune its rate of docking by adjusting $\lambda_{B}$ and $p_{0}$. 

In general, the bending simulations reach higher $M$ values than planar BCPM simulations run with the same parameters. (For example, the parameters used in Region I of Fig.~\ref{fig:Bending_Regions}(b) would be expected to yield only $M\approx0.8$ on average in a planar BCPM scenario, according to Fig.~\ref{fig:Jamming_and_regime_map}(a).) This is possibly due to our advancement criterion giving rise to a ``pre-matched'' boundary condition for the region II subsystem that aids matching, as well as apparently less favorable conditions for domain walls in the bending simulations (a few are observed to form, but resolve quickly).


\section{\label{sec:level1}DISCUSSION }


In this work we introduce a minimal model of epithelial docking that couples two 2D area- and perimeter-elasticity models via cell edges, where the relevant bilayer adhesion molecules are assumed to be concentrated. Our model is a first step towards a quantitative understanding of the role of cell edge matching in a variety of tissue fusion processes, a role which has evaded conclusive experimental observation due to difficulties in high-resolution imaging within or parallel to the fusion plane. 
Future experiments could test the validity of our model, and particularly the form of the bilayer coupling term (Eq.~\eqref{eq:Coupling_Hamiltonian}), through comparison with our predictions for parameter values that maximize edge matching, and whether coordinated T1 events and domain walls are observed.

A key finding of this work is the existence of a region in the $p_0$-$\lambda_B$ plane where $\langle M_f\rangle\approx1$, indicative of successful epithelial docking (Fig.~\ref{fig:Jamming_and_regime_map}(a)). The range of values $\lambda_B\approx$ 6-8 in this region represents a balance between in-plane and out-of-plane energy scales as illustrated by the following argument: Consider an ``armchair''-shaped interface (i.e., the shape of the right or left boundary in Fig.~\ref{fig:Edgematch_definition}) between two cells in layer 1 that are both at their target area and perimeter. The highest-energy, non-fragmentary, single-pixel perturbation of this interface changes each cell's area by $\pm a$, and changes each cell's perimeter by $4s$. According to Eq.~\eqref{eq:Monolayer_Hamiltonian}, the perturbation energy is $2\lambda_Aa^2+2\lambda_P(4s)^2 = 14.3$ in simulation units. If this perturbation gains length $s$ worth of cell edge match with layer 2, the bilayer energy is reduced by $2\lambda_B$, and the pertubation becomes energetically favorable when $0=14.3-2\lambda_B$, or $\lambda_B\approx7$. However, if $\lambda_B$ is too far in excess of this value, the system tends to get stuck in partially matched metastable states with high energy barriers preventing the intra-layer rearrangements that would be required to achieve $\langle M_f\rangle\approx1$. That successful docking also requires the system to be in the fluid-like regime ($\langle P_\sigma/\sqrt{A_\sigma}\rangle\gtrsim4.6$) is not surprising, given the many intra-layer rearrangements and coordinated T1 transitions that need to take place to match the two layers. What is surprising, and warranting further investigation, is the observation that introducing a small $\lambda_B$ appears to shift the solid-like to fluid-like transition to smaller $p_0$ (Fig.~\ref{fig:Jamming_and_regime_map}(d)), reminiscent of the effect of activity~\cite{bi2016motility}. We speculate that in this limit, where the system never achieves more than $\sim50\%$ matching, Eq.~\eqref{eq:Coupling_Hamiltonian} may in fact act like a source of activity by subjecting the cells in one layer to random forcing from the other layer. 

Our other key finding is the existence of domain wall-like defects separating completely matched regions of the bilayer system from each other. Such defects have no counterpart in purely 2D cellular models. Whether or not these structures exist in real biological systems is, as far as we know, an open question. It is interesting to speculate on what the biological consequences might be, e.g., would a domain wall be a source of mechanical weakness that might facilitate bending of the bilayer, could it provide a pathway for vasculature or some other kind of transport, etc. The crude bending extension we have studied here suggests domain walls are less relevant to the docking of curved epithelial surfaces in which the fusion front propagates more unidirectionally. Future work will also include a more realistic treatment of curvature and bending in a bilayer docking model, exploring this question and others.


\section{\label{sec:level1} ACKNOWLEDGMENTS }


T.S., A.J., and T.A.E acknowledge funding from the National Eye Institute of the National Institutes of Health under Award Number R15EY035473. The content is solely the responsibility of the authors and does not necessarily represent the official views of the National Institutes of Health. This work utilized the Alpine High-Performance Computing resource at the University of Colorado Boulder. Alpine is jointly funded by the University of Colorado Boulder, the University of Colorado Anschutz, Colorado State University, and the National Science Foundation (award 2201538). Conversations with Jen Schwarz, Daniel Sussman, Jessica Feldman, Lauren Cote, Dinah Loerke, and Todd Blankenship were valuable in this work. Preliminary research by undergraduate students Indigo Peppard and Micah Garcia helped motivate this study.   


\section{\label{sec:level1} DATA AVAILABILITY}


The code used in this paper is visible at \url{https://github.com/TroyS/Bilayer_Cellular_Potts/tree/main} .


\section{appendix}


\subsection{Displacement and Diffusion Coefficients }

\begin{figure}[h]
\centering
\includegraphics[width=\linewidth]{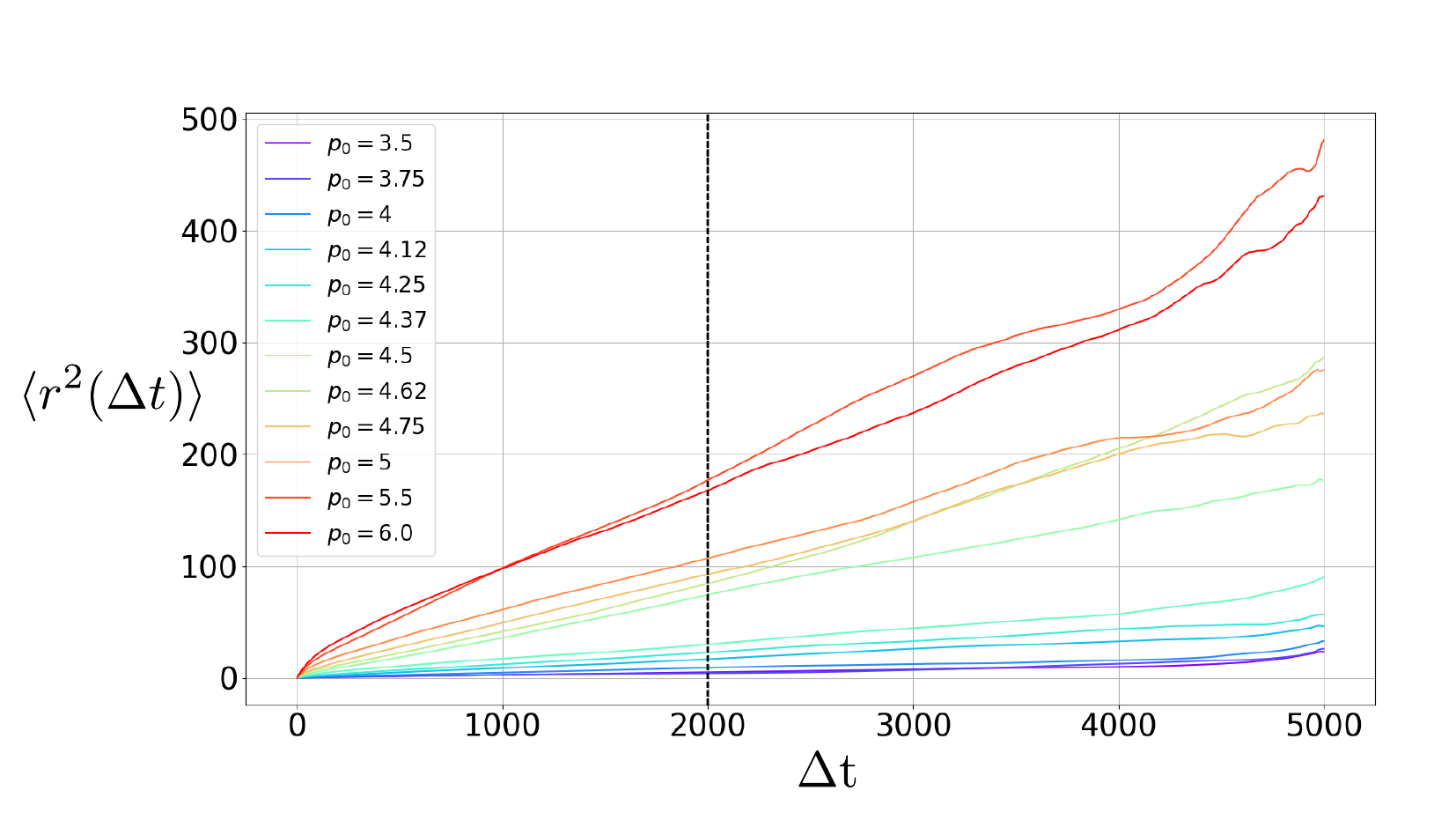}
\caption{Cell center MSD vs time curves for various $p_{0}$ when $\lambda_{B} = 0$. Diffusion coefficients are calculated by taking the slope of an MSD curve in the range $0 \le t \le 2000$.   }
\label{fig:MSD_curves}
\end{figure}

In a system with a large number of moving particles, particle motion is typically characterized by the diffusion coefficient. Given the positions of various particles over time $\vec{r}_{i}(t)$, the traditional method for computing the diffusion coefficient $D$ is finding the slope of an MSD (mean squared displacement) versus time plot,
\begin{equation}
\langle \left[\vec {r}(t)-\vec{r}(0)\right]^2 \rangle = 2nDt
\end{equation}
where $\langle \left[\vec{r}(t) - \vec{r}(0)\right]^2 \rangle$ denotes an ensemble average over all the particles in the system and $n$ is the number of dimensions. When tracking the motion of a single particle, the ensemble average is replaced by an average over dimensionless time lags $\Delta t$ ~\cite{Michalet2011,DALCASTEL2025130524},
\begin{eqnarray}
\vec{r}^2(\Delta t) = \frac{1}{N-\Delta t} \sum_{t=0}^{N-\Delta t}\left[\vec{r}(t+\Delta t)-\vec{r}(t)\right]^2 , \\ \nonumber
\Delta t = 1, ..., N-1
\end{eqnarray}
with equality to the ensemble average holding for an ergodic system. In practice this definition introduces errors \cite{bullerjahn2020optimal} due to fewer time intervals included in the averaging at higher $\Delta t$ and correlations in particle positions between times in a computational simulation. With periodic boundary conditions used in our simulations the position over time data $\vec{r}(t)$ of some of the cells spikes when those cells cross a periodic boundary, introducing another source of error. The diffusion coefficients shown in \ref{fig:Jamming_and_regime_map} (a) were obtained by calculating the time averaged MSD for every cell in a simulation, averaging the MSDs over cells that did not have spikes, and finding the slope of the MSD vs time plot in the region $0 \le \Delta t \le 2000 $ as shown in Fig.~\ref{fig:MSD_curves}.

\begin{figure}[h]
\centering
\includegraphics[width=0.7\linewidth]{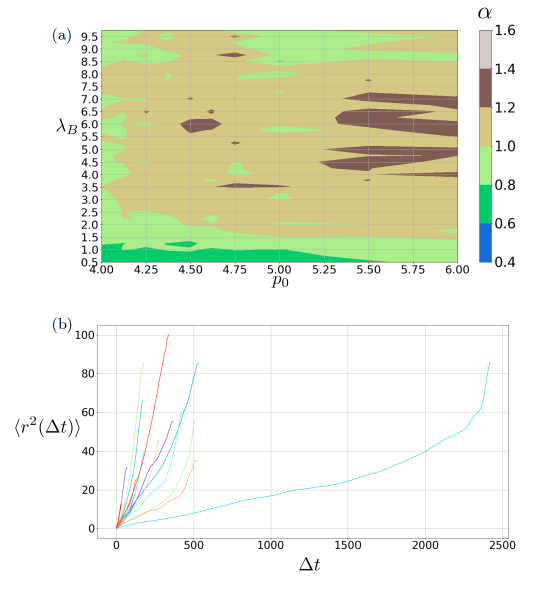}
\caption{(a) A regime map of scaling exponent $\alpha$ for MSDs at the characteristic time. (b) Sample MSD curves for different seeds in the $p_{0}=5, \lambda_{B}=4.5$ condition, illustrating why we use a cutoff of $\tau/2$ to avoid the superdiffusive regime when computing diffusion coefficients.  }
\label{fig:Regime_map_diff_coeff}
\end{figure}

 Computing diffusion coefficients for the $\lambda_{B}\ne 0$ up until the characteristic time is more error prone than in the $\lambda_{B}=0$ case due to the varying characteristic times in each simulation making the linear regime in each MSD curve harder to identify and the fewer points used in simulations with short characteristic times increasing the error in diffusion coefficients. Anomalous diffusion possibly occurs in this regime, defined with $MSD(t) = K_{\alpha}t^{\alpha}$ where $K_{\alpha}$ is a generalized diffusion coefficient and $\alpha$ defines the diffusive regime, Brownian motion corresponding to $\alpha=1$.
 A regime map of $\alpha$ for the computed MSDs at characteristic time for each condition and sample MSD curves for different seeds in the $p_{0}=5, \lambda_{B} = 4.5$ condition are shown in Fig.~\ref{fig:Regime_map_diff_coeff} (a) and (b) respectively.

The mean distance traveled at time $t$ used to quantify cell mobility is defined  
\begin{equation}
X(t) = \frac{1}{N}\sum_{i=1}^{N}  \sum_{j=1}^{t} | X_{i}(j)-X_{i}(j-1) |
\end{equation}
where $X_{i}$ represents either the $x$ or $y$ component of the position vector $\vec{r_{i}}(t)$. The components of the speed shown in Fig.~\ref{fig:Jamming_and_regime_map} (c) are given by $\frac{X(\tau)}{\tau}$ where we exclude cells that cross periodic boundaries.

\subsection{Neighbor Changes }

\begin{figure}[h]
\centering
\includegraphics[width=0.5\linewidth]{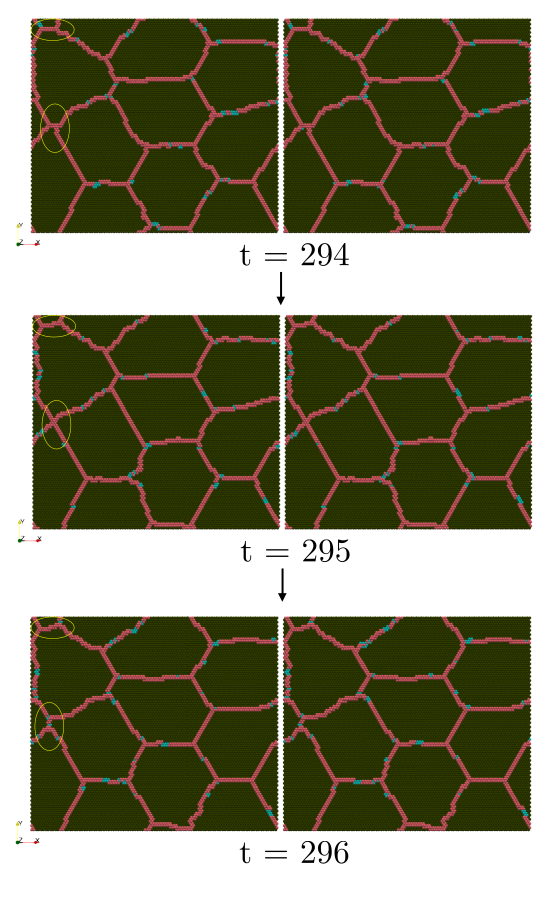}
\caption{An example of an event with two neighbor changes, where two cells that are disconnected via a T1 transition remain connected via the periodic boundary conditions.   }
\label{fig:Neighbor_change_of_two}
\end{figure}

As discussed in the main text, cell rearrangement events involving less than 4 neighbor changes are possible in the BCPM for various reasons, with an example shown in Fig.~\ref{fig:Neighbor_change_of_two}.

\providecommand{\noopsort}[1]{}\providecommand{\singleletter}[1]{#1}%

\end{document}